\documentclass[prd,amsmath,amssymb,superscriptaddress,nofootinbib,twocolumn]{revtex4}
\usepackage{amsfonts}
\usepackage{multirow}
\usepackage{makecell}
\usepackage{mathrsfs}
\usepackage{graphicx}
\usepackage{amsmath}
\usepackage{amssymb}
\usepackage{bm}
\usepackage{bbm}
\usepackage{color}
\usepackage{ulem}
\usepackage{array}
\usepackage{diagbox}
\usepackage{float}

\newcommand{\nn}{\nonumber}

\newcommand{\beq}{\begin{equation}}
\newcommand{\eeq}{\end{equation}}
\newcommand{\bqa}{\begin{eqnarray}}
\newcommand{\eqa}{\end{eqnarray}}

\newcommand{\bseq}{\begin{subequations}}
\newcommand{\eseq}{\end{subequations}}

\makeatletter

\begin{document}

\title{ Next-to-next-to-leading-order QCD corrections to pion electromagnetic form factors
}
\author{Long-Bin Chen~\footnote{chenlb@gzhu.edu.cn}}
\affiliation{ School of Physics and Materials Science, Guangzhou University, Guangzhou 510006, China \vspace{0.2cm}}
\author{Wen Chen~\footnote{chenwenphy@gmail.com}}
\affiliation{Institute of Quantum Matter, South China Normal University, Guangzhou 510006, China}
\affiliation{Zhejiang Institute of Modern Physics, School of Physics, Zhejiang University, Hangzhou, Zhejiang
310027, China\vspace{0.2cm}}
\author{Feng Feng~\footnote{f.feng@outlook.com}}
\affiliation{China University of Mining and Technology, Beijing 100083, China\vspace{0.2cm}}
\affiliation{Institute of High Energy Physics and Theoretical Physics Center for Science Facilities, Chinese Academy of Sciences, Beijing 100049, China\vspace{0.2cm}}
\author{Yu Jia~\footnote{jiay@ihep.ac.cn}}
\affiliation{Institute of High Energy Physics and Theoretical Physics Center for Science Facilities, Chinese Academy of Sciences, Beijing 100049, China\vspace{0.2cm}}
\affiliation{School of Physical Sciences,
University of Chinese Academy of Sciences, Beijing 100049, China\vspace{0.2cm}}
\date{\today}

\begin{abstract}
We calculate the next-to-next-to-leading order (NNLO) QCD radiative correction to the pion electromagnetic form factor with large momentum transfer.
We explicitly verify the validity of the collinear factorization to two-loop order for this observable, and obtain
the respective IR-finite two-loop hard-scattering kernel in the closed form.
The NNLO QCD correction turns to be positive and significant. Incorporating this new ingredient of correction,
we then make a comprehensive comparison between the finest theoretical predictions and numerous data for both space-like and time-like pion form factors.
Our phenomenological analysis provides strong constraint on the second Gegenbauer moment of the pion light-cone distribution amplitude (LCDA)
obtained from recent lattice QCD studies.
\end{abstract}

\maketitle

\noindent{\color{blue}\it Introduction.}
Originally proposed by Yukawa as the strong nuclear force carrier in 1935~\cite{Yukawa:1935xg},
subsequently discovered in the cosmic rays in 1947~\cite{Lattes:1947mw},
the $\pi$ mesons have always occupied the central stage throughout the historic advancement of the strong interaction.
As the lightest particles in the hadronic world (hence the highly-relativistic bound systems composed of light quark and gluons),
$\pi$ mesons entail extremely rich QCD dynamics, exemplified by the color confinement and chiral symmetry breaking.
Notwithstanding extensive explorations during the past decades, there still remain some great myths about
the internal structure of the $\pi$ mesons.

 A classic example of probing the internal structure of the charged $\pi$ is the pion electromagnetic
 (EM) form factor:
\beq
\langle \pi^+(P') \vert J_{\rm em}^\mu \vert \pi^+(P)\rangle = F_\pi(Q^2) (P^\mu +P'^\mu),
\eeq
with $Q^2\equiv -(P-P^\prime)^2$. The electromagnetic current is defined by $J^\mu_{\rm em}=\sum_f e_f \bar{f} \gamma^\mu f$,
with $e_u=2/3$ and $e_d=-1/3$ indicating the electric charges of the $u$ and $d$ quarks.

During the past half century, the pion EM form factor has been intensively studied experimentally~\cite{Bebek:1974ww,Bebek:1976qm,Bebek:1977pe,Ackermann:1977rp,Brauel:1979zk,Dally:1981ur,Dally:1982zk,Amendolia:1984nz,
NA7:1986vav,JeffersonLabFpi:2000nlc,JeffersonLabFpi:2007vir,JeffersonLabFpi-2:2006ysh,JeffersonLab:2008gyl,JeffersonLab:2008jve,
CMD-2:2001ski,CMD-2:2003gqi,Achasov:2005rg,Achasov:2006vp,Aulchenko:2006dxz,CMD-2:2006gxt,KLOE:2008fmq,BaBar:2009wpw,KLOE:2010qei,KLOE:2012anl,BaBar:2012bdw,KLOE-2:2017fda,BESIII:2015equ}.
From the theoretical perspective, the pion EM form factor at small $Q^2$ can be investigated in chiral perturbation theory~\cite{Gasser:1983yg}
and lattice QCD~\cite{Martinelli:1987bh,Draper:1988bp,Feng:2019geu,Wang:2020nbf,Gao:2021xsm}.
On the other hand, at large momentum transfer, the $F_\pi(Q^2)$ is expected to
be adequately described by perturbative QCD.
Within the collinear factorization framework tailored for hard exclusive reactions~\cite{Lepage:1979zb,Lepage:1979za,Lepage:1980fj,Efremov:1978rn,Efremov:1979qk,Duncan:1979ny,Duncan:1979hi}
(for a review, see \cite{Chernyak:1983ej}), at the lowest order in $1/Q$,
the pion EM form factor can be expressed in the following form:
\beq
F_\pi(Q^2) = \int\!\!\! \int dx\,dy\,\Phi_\pi^*(x,\mu_F) T(x,y,\frac{\mu_R^2}{Q^2},\frac{\mu_F^2}{Q^2}) \Phi_\pi(y,\mu_F),
\label{EMFF:collinear:factorization}
\eeq
where $T(x,y)$ signifies the perturbatively calculable hard-scattering kernel, and $\Phi_\pi(x,\mu_F)$
represents the nonperturbative yet universal leading-twist pion light-cone distribution amplitude (LCDA), $i.e.$,
the probability amplitude of finding the valence $u$ and $\bar{d}$ quark inside $\pi^+$
carrying the fractional momenta $x$ and $\bar{x}\equiv 1-x$, respectively.
The leading-twist pion LCDA assumes the following operator definition:
\bqa
\Phi_\pi(x,\mu_F) &=& \int \frac{d z^-}{2 \pi i} e^{i z^- x P^+}\left\langle0\left|\bar{d}(0) \gamma^+ \gamma_5 \right. \right.
\nn \\
 && \times \left.\left. {\cal W}(0, z^-) u(z^-)\right| \pi^+(P)\right\rangle,
\label{pi:LCDA:operator:def}
\eqa
with ${\cal W}$ signifies the light-like gauge link to ensure the gauge invariance.
Conducting the UV renormalization for \eqref{pi:LCDA:operator:def},
one is led to the celebrated Efremov-Radyushkin-Brodsky-Lepage (ERBL) evolution equation~\cite{Lepage:1980fj, Efremov:1979qk}:
\begin{equation}
\frac{d\Phi_\pi(x,\mu_F)}{d\ln\mu^2_F} = \int_0^1\!\! dy \, V(x,y) \, \Phi_\pi(y,\mu_F),
\label{ERBL:evolution:eq}
\end{equation}
with $V(x,y)$ referring to the perturbatively calculable ERBL kernel.

Eq.~\eqref{EMFF:collinear:factorization} is expected to hold to all orders in perturbative expansion.
The hard-scattering kernel can thus be expanded in the power series:
\beq
T  = \frac{16C_F\pi\alpha_s}{ Q^2} \left\{ T^{(0)} + {\alpha_s\over \pi} T^{(1)} +
\left({\alpha_s\over \pi}\right)^2 T^{(2)} + \cdots \right\},
\eeq
with $C_F={N_c^2-1\over 2N_c}$, and $N_c=3$ is the number of colors.

The leading order (LO)
result was known shortly after the advent of QCD~\cite{Lepage:1979za,Efremov:1979qk,Duncan:1979hi,Lepage:1980fj,Chernyak:1977fk,Farrar:1979aw,Chernyak:1980dj}.
The next-to-leading order (NLO) correction was originally computed by three groups in early 80s~\cite{Field:1981wx,Dittes:1981aw,Sarmadi:1982yg}.
Scrutinizing the previous calculations, in 1987 Braaten and Tse traced the origin of the discrepancies among the earlier
work and presented the correct expression of the NLO hard-scattering kernel~\cite{Braaten:1987yy}.
In 1998,  Meli\'{c}, Ni\u{z}i\'{c}, and Passek conducted a comprehensive phenomenological study by incorporating the NLO correction as well as the evolution effect of pion LCDA~\cite{Melic:1998qr}. The central goal of this work is to compute the next-to-next-to-leading order (NNLO) perturbative correction to pion EM form factor,
and critically examine its phenomenological impact~\footnote{It is worth mentioning that, historically there has been debate on the applicability of collinear factorization
at experimentally accessible moderate energy range, say, $Q<6$ GeV~\cite{Isgur:1989cy,Bolz:1996sw,Radyushkin:1998rt,Braun:1999uj}.
The central issue is whether the nonfactorizable soft mechanism can dominate the perturbative QCD contribution or not.}.

\vspace{0.1 cm}

\noindent{\color{blue}\it Setup of perturbative matching.}
The strategy of deducing the short-distance coefficients is through the standard matching procedure.
Since the hard-scattering kernel is insensitive to the long-distance physics, it is legitimate to replace the physical $\pi^+$
by a free quark-antiquark pair $u \bar{d}$, and compute both sides of
\eqref{EMFF:collinear:factorization} in perturbation theory, order by order in $\alpha_s$.
To make things simpler, we neglect the transverse motion and assign the momenta of the $u$ and $\bar{d}$ in the incoming ``pion" to be
$u P$ and $\bar{u} P$, and assign the momenta of the $u$ and $\bar{d}$ in the outgoing ``pion" to be
$v P^\prime$ and $\bar{v} P^\prime$, with $u, v$ ranging from $0$ to $1$.

\begin{figure}
\centering
\includegraphics[width=0.5\textwidth]{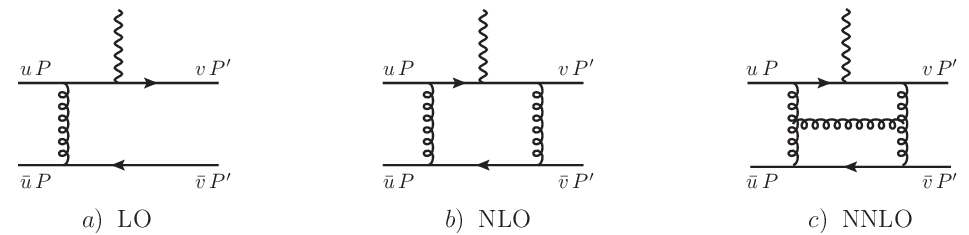}
\caption{Sample parton-level Feynman diagrams for the reaction $\gamma^*\pi(P)\to\pi(P^\prime)$ at various perturbative orders.}
\label{fig:feynman-diagram}
\end{figure}

In the left-hand side of \eqref{EMFF:collinear:factorization}, we extract the scalar form factor $F(u,v)$ through the
partonic reaction $\gamma^*+u(uP) \bar{d}({\bar{u}}P)\to u(v P') \bar{d}({\bar{v}}P')$. Some typical Feynman diagrams through two-loop order
are depicted in Fig.~\ref{fig:feynman-diagram}. It is subject to a perturbative expansion:
\beq
\begin{split}
F(u,v) = &\frac{16C_F\pi\alpha_s}{ Q^2}\left[F^{(0)}(u,v) + {\alpha_s\over \pi} F^{(1)}(u,v)\right.\\
&\left. +
\left({\alpha_s\over \pi}\right)^2 F^{(2)}(u,v) + \cdots\right].
\label{fictitious:pion:FF:expansion}
\end{split}
\eeq

In the right-hand side of \eqref{EMFF:collinear:factorization},
one can expand the renormalized ``pion" LCDA as
\beq
\Phi(x|u)  = \Phi^{(0)}(x|u) + {\alpha_s\over \pi} \Phi^{(1)}(x|u) +
\left({\alpha_s\over \pi}\right)^2 \Phi^{(2)}(x|u) + \cdots.
\label{fictitious:pion:DA:expansion}
\eeq

At tree level, the fictitious pion LCDA in \eqref{pi:LCDA:operator:def} simply reduces to
$\Phi^{(0)}(x|u) = \delta(x-u)$~(up to a normalization factor which also appears in $F(u,v)$).
By equating both sides of \eqref{EMFF:collinear:factorization}, one reproduces the well-known tree-level expression $T^{(0)}(x,y)$~\cite{Lepage:1979zb,Lepage:1979za,Lepage:1980fj,Efremov:1978rn,Efremov:1979qk,Duncan:1979ny,Duncan:1979hi}
\begin{equation}
T^{(0)}(x,y) =  \frac{e_u}{\bar{x}\bar{y}}(1-\epsilon) -
\genfrac[]{0pt}{0}{e_u \to e_d}{\bar{x}\to x, \bar{y}\to y},
\end{equation}
which holds true in $d=4-2\epsilon$ spacetime dimension.

Once beyond the tree level, the UV and IR divergences inevitably arise and we use the dimensional regularization (DR) to regularize both types of divergences.
Nevertheless, the bare ``pion" LCDA remains intact since the scaleless integrals vanish in DR.
The renormalized ``pion" LCDA is related to the bare one via
\beq
\Phi(x|u) = \int\!\! dy Z(x,y) \Phi_{\rm bare}(y|u) = Z(x,u),
\label{LCDA:renormalization}
\eeq
which is solely comprised of various IR poles.

$Z(x,y)$ in \eqref{LCDA:renormalization} signifies the renormalization function in the $\overline{\rm MS}$ scheme,
which can be cast into the following Laurent-expanded form in $\epsilon$:
\begin{equation}
    Z(x,y) = \delta(x-y) + \sum_{k=1}^\infty \frac{1}{\epsilon^k} Z_k(x,y),
\label{Z:renormalization:function}
\end{equation}

Note that the prefactor of single pole in \eqref{Z:renormalization:function} is related to the ERBL kernel
$V(x,y)$ in \eqref{ERBL:evolution:eq} via $V(x,y) = -\alpha_s \partial Z_1/\partial \alpha_s$~\cite{Floratos:1977au}.
Note that the two-loop~\cite{Sarmadi:1982yg, Dittes:1983dy, Katz:1984gf, Mikhailov:1984ii, Belitsky:1999gu}
and three-loop corrections~\cite{Braun:2017cih} to $V(x,y)$ have been available.

The two-loop renormalized ``pion" LCDA $\Phi^{(2)}$ also contains double IR pole.
The $Z_2$ can be obtained through the recursive relation~\cite{Becher:2005pd}
\beq
\alpha_s \frac{\partial Z_{2}}{\partial\alpha_s} = \alpha_s \frac{\partial Z_{1}}{\partial\alpha_s} \otimes Z_{1} + \beta(\alpha_s) \frac{\partial Z_1}{\partial\alpha_s},
\eeq
where $d\alpha_s/d\ln\mu^2 = - \epsilon \alpha_s+ \beta(\alpha_s) $.

With the aid of \eqref{LCDA:renormalization} and \eqref{Z:renormalization:function}, we then determine the ${\cal O}(\alpha_s)$ and ${\cal O}(\alpha^2_s)$
corrections to the renormalized ``pion" LCDA in \eqref{fictitious:pion:DA:expansion}.

At one-loop order, the matching equation for a fictitious pion state becomes
\begin{eqnarray}
Q^2 F^{(1)}(u,v) &=& T^{(1)}(u,v) + \Phi^{(1)}(x|u) \mathop{\otimes}\limits_{x} T^{(0)}(x,v)
\nn\\
&& + \Phi^{(1)}(y|v) \mathop{\otimes}\limits_{y} T^{(0)}(u,y),
\eqa
where $\mathop{\otimes}\limits_{x}$ signifies the convolution over $x$.
Note that the renormalized scalar form factor $F^{(1)}(u,v)$ still contains single collinear pole. However, the renormalized $\Phi^{(1)}(x|u)$ and $\Phi^{(1)}(y|v)$
also contains the same IR poles. Upon solving this matching equation, one ends with both UV and IR-finite $T^{(1)}(x,y)$.
Our expressions agree with the known NLO result~\cite{Melic:1998qr}.

To the desired two-loop order, the following matching equation descends from \eqref{EMFF:collinear:factorization}:
\bqa
 && Q^2F^{(2)}(u,v) = T^{(2)}(u,v) + \Phi^{(2)}(x|u) \mathop{\otimes}\limits_{x} T^{(0)}(x,v)
\nn\\
&& + \Phi^{(2)}(y|v) \mathop{\otimes}\limits_{y} T^{(0)}(u,y) + \Phi^{(1)}(x|u) \mathop{\otimes}\limits_{x} T^{(1)}(x,v)
\nn
\\
&& + \Phi^{(1)}(y|v) \mathop{\otimes}\limits_{y} T^{(1)}(u,y)
\label{two:loop:matching:equation}
\\
&& + \Phi^{(1)}(x|u) \mathop{\otimes}\limits_{x} T^{(0)}(x,y) \mathop{\otimes}\limits_{y} \Phi^{(1)}(y|v), \nonumber
\eqa
More Severe IR divergences are expected to arise in both $F^{(2)}(u,v)$ and $\Phi^{(2)}(x|u)$.
Clearly one also needs compute $T^{(1)}(x,y)$ to ${\cal O}(\epsilon)$.

\vspace{0.1 cm}

\noindent{\color{blue}\it Description of the calculation.}
We use {\tt HepLib}~\cite{Feng:2021kha} and {\tt FeynArts}~\cite{Hahn:2000kx} to generate Feynman diagrams and the corresponding amplitudes
for the reaction $\gamma^*+u(uP) \bar{d}({\bar{u}}P)\to u(v P')\bar{d}({\bar{v}}P')$.
We employ the covariant projector technique to enforce each $u\bar{d}$ pair to bear zero helicity.
For our purpose it suffices to adopting the naive anticommutation relation to handle $\gamma_5$ in DR.
There are about 1600 two-loop diagrams, one of which is depicted in Fig.~\ref{fig:feynman-diagram}$c)$.
We employ the package {\tt Apart}~\cite{Feng:2012iq} to conduct partial fraction, and {\tt FIRE}~\cite{Smirnov:2019qkx} for integration-by-part
reduction. We end up with 116 independent master integrals (MIs).
The MIs are calculated by utilizing the differential equations method~\cite{Kotikov:1990kg,Remiddi:1997ny,Henn:2013pwa}.
Note that these MIs are considerably more involved than than what are encountered in the two-loop corrections for the $\pi-\gamma$ transition form factor~\cite{Gao:2021iqq,Braun:2021grd}.
We have attempted two independent ways to construct and solve the differential-equation systems, one of which is based on the method developed in \cite{Chen:2019mqc,Chen:2019fzm,Chen:2020wsh,Chen:2023hmk}.
The analytic results are expressed in terms of the Goncharov Polylogarithms (GPLs)~\cite{Goncharov:1998kja}.
Two independent calculations yield the identical answer. We also numerically check our results against the package
\texttt{AMFLOW}~\cite{Liu:2022chg} and found perfect agreement.
Technical details will be included in the future long write-up.

Upon renormalizing the QCD coupling in $\overline{\rm MS}$ scheme, we end up with a rather lengthy expression for $F^{(2)}(u,v)$.
Being UV finite, it still contains severe IR divergences which start at order-$1/\epsilon^2_{\rm IR}$. Inspecting the matching equation
\eqref{two:loop:matching:equation}, piecing all the known ingredients together,
we are able to solve for the intended two-loop hard-scattering kernel.
Hearteningly, $T^{(2)}(x,y)$ is indeed IR finite. Therefore, our explicit calculation verifies that
the collinear factorization does hold at two-loop level for the pion EM form factor.
The analytical expression of $T^{(2)}(x,y)$ is too lengthy to be reproduced here.
For the sake of clarity, in the suppletory material we provide the asymptotic expressions of $T^{(1,2)}(x,y)$
near the endpoint regions.

\vspace{0.1 cm}

\noindent{\color{blue}\it Master formula for pion EM form factor at NNLO.}
Given a certain parametrized form of pion LCDA, the two-fold integration in \eqref{EMFF:collinear:factorization}
turns out to be difficult to conduct numerically, mainly due to numerical instability caused by the spurious singularity as $x\to y$/$x\to \bar{y}$ in
$T^{(2)}(x, y)$. Our recipe to circumvent this technical challenge is to present the two-loop results in an analytical manner,
which enables us to achieve exquisite precision.

The leading-twist pion LCDA is conveniently expanded in the Gegenbauer polynomial basis:
\begin{subequations}
\begin{align}
\Phi_\pi(x,\mu_F) = \frac{f_\pi}{2\sqrt{2N_c}} {\sum_{n = 0}}' a_n(\mu_F) \psi_n(x),
\\
\psi_n(x) = 6x\bar{x} C_n^{3/2}(2x-1),
\end{align}
\label{pion:DA:expanded:in:Gegenbauer}
\end{subequations}
where the pion decay constant $f_\pi = 0.131$ GeV, and $\sum'$ signifies the sum over even integers.
Note all the nonperturbative dynamics is encoded in the Gegenbauer moments $a_n(\mu_F)$.

Substituting \eqref{pion:DA:expanded:in:Gegenbauer} into \eqref{EMFF:collinear:factorization},
we reexpress the pion EM form factor as
\bqa
& &Q^2 F_\pi(Q^2)= \frac{2C_F \pi^2 (e_u-e_d)f_\pi^2}{3}\times \nonumber\\
& & \sum_{k=0} \left(\alpha_s\over \pi\right)^{k+1}  {\sum_{m,n}}' a_n(\mu_F) a_m(\mu_F) {\cal T}^{(k)}_{mn},
\label{Master:formula:pion:EM:FF}
\eqa
with ${\cal T}^{(k)}_{mn}$  defined by
\begin{equation}
    {\cal T}^{(k)}_{mn} =\frac{1}{e_u-e_d} \psi_m(x) \mathop{\otimes}\limits_{x} T^{(k)}
    \left(x,y,\frac{\mu_R^2}{Q^2},\frac{\mu_F^2}{Q^2}\right) \mathop{\otimes}\limits_{y} \psi_n(y).
\label{T:k:mn}
\end{equation}
For simplicity, we will set $\mu_R=\mu_F=\mu$ and $n_L=3$ from now on.
The two-fold integrations in (\ref{T:k:mn}) can be readily worked out at tree and one-loop levels.
For instance, we have
\begin{align}
{\cal T}^{(0)}_{mn} = 9,\qquad {\cal T}^{(1)}_{00} = \frac{1}{4}  (81L_\mu+237),
\end{align}
with $L_\mu \equiv \ln({\mu^2}/{Q^2})$.

Remarkably, the two-loop coefficients ${\cal T}^{(2)}_{mn}$ can also be computed analytically, thanks to the fact that
$T^{(2)}$ can be expressed in terms of the GPLs.
Although the integrand in \eqref{T:k:mn} contains about ${\cal O}(10^5)$ individual terms, the final result after two-fold integration becomes exceedingly
compact, which can be expressed in terms of the rational numbers and Riemann zeta function.
For instance, the expression of ${\cal T}^{(2)}_{00}$ reads
\begin{widetext}
\begin{equation}
{\cal T}^{(2)}_{00} =  \frac{729 L_{\mu }^2}{16}-(8 \zeta_3+\frac{35 \pi ^2 }{6}-\frac{2961
  }{8}) L_{\mu }+205 \zeta_5-\frac{3 \pi ^4}{20}-\frac{651
   \zeta_3}{2}-\frac{275 \pi
   ^2}{24}+821 .
\end{equation}
\end{widetext}
Due to the length restriction, we refrain from providing the analytical expressions for other ${\cal T}^{(1,2)}_{mn}$. For reader's
convenience, in Table~\ref{tab:Tmn-numeric-values} we tabulate the numerical values of ${\cal T}^{(1,2)}_{mn}$ for $0\le m,n\le 6$,
which is sufficient for most phenomenological analysis.

\begin{table}[ht]
\begin{tabular}{c|cc|ccc}
\hline\hline
(m,n) & $c_1$ & $c_2$ & $d_1$ & $d_2$ & $d_3$ \\
\hline\hline
(0,0) & 20.25 & 59.25 & 45.5625& 302.936& 514.600 \\
(0,2) & 32.75 & 112.473 & 96.4306& 735.637& 1498.75 \\
(0,4) & 38.45 & 147.638 & 125.390& 1049.88& 2346.49 \\
(0,6) & 42.2571 & 174.359 & 146.743& 1306.76& 3103.33 \\
(2,2) & 45.25 & 192.871 & 164.660& 1513.95& 3690.60 \\
(2,4) & 50.95 & 240.181 & 201.536& 2024.03& 5371.58 \\
(2,6) & 54.7571 & 274.974 & 228.176& 2424.45& 6796.36 \\
(4,4) & 56.65 & 292.970 & 242.021& 2640.85& 7589.17 \\
(4,6) & 60.4571 & 331.411 & 271.074& 3118.35& 9432.36 \\
(6,6) & 64.2643 & 372.282 & 301.736& 3651.19& 11598.1 \\
\hline\hline
\end{tabular}
\caption{The numerical values for ${\cal T}^{(1)}_{mn} = c_1 L_\mu + c_2$ and ${\cal T}^{(2)}_{mn} = d_1 L_\mu^2 + d_2 L_\mu+d_3$,
with $0\le m,n\le 6$. }
\label{tab:Tmn-numeric-values}
\end{table}

With the input from Table~\ref{tab:Tmn-numeric-values}, Eq.~\eqref{Master:formula:pion:EM:FF} constitutes our master formula for yielding
phenomenological predictions through the two-loop accuracy.
Compared with the original factorization formula \eqref{EMFF:collinear:factorization}, we have simplified an integration task into an algebraic one.

It is straightforward to adapt our master formula from the space-like region to the time-like one,
provided that one makes the replacement $L_\mu \to  L_\mu+ i\pi$ in Table~\ref{tab:Tmn-numeric-values},
with $Q^2$ now indicating the squared invariant mass of the $\pi^+\pi^-$ pair.

\vspace{0.1 cm}

\noindent{\color{blue}\it Input parameters.}
As the key nonperturative input, our knowledge on the pion LCDA is still not confirmative enough.
In the early days, it is popular to assume asymptotic form, CZ parametrization~\cite{Chernyak:1983ej} and the BSM parameterizations~\cite{Mikhailov:2016klg,Bakulev:2001pa}.
In recent years there have emerged extensive investigations on the profile of the pion DA from different methodologies, including QCD light-cone sum rule~\cite{Ball:2006wn} with nonlocal
condensate~\cite{Mikhailov:2016klg,Stefanis:2020rnd} or fitted from dispersion relation~\cite{Cheng:2020vwr} or Platykurtic~\cite{Stefanis:2014nla},
Dyson-Schwinger equation~\cite{Chang:2013pq,Raya:2015gva}, basis light-front quantization~\cite{Lan:2019rba},
light-front quark model~\cite{Choi:2014ifm}, holographic QCD~\cite{Chang:2016ouf}, and very recently, from the lattice simulation \cite{RQCD:2019osh,LatticeParton:2022zqc}.
The predicted values of various Gegenbauer moments are scattered in a wide range.

Since lattice QCD provides the first-principle predictions, in this Letter we will take the most recent lattice results as inputs.
In 2019 \texttt{RQCD} Collaboration has presented a precise prediction for the second Gegenbauer moment of pion LCDA in $\overline{\rm MS}$ scheme, with $a_2(2\;{\rm GeV})=0.116^{+0.019}_{-0.020}$~\cite{RQCD:2019osh}.

An important progress in lattice QCD is expedited by the advent of the Large-momentum effective theory (LaMET) a decade ago~\cite{Ji:2013dva,Ji:2014gla},
which allows one to access the light-cone distributions in Euclidean lattice directly in the $x$ space. Very recently, the
\texttt{LPC} Collaboration has presented the whole profile of the pion LCDA~\cite{LatticeParton:2022zqc}, from which various Gegenbauer moments can be inferred:
$a_2(2\;{\rm GeV})=0.258\pm0.087$, $a_4(2\;{\rm GeV})=0.122\pm0.056$, $a_6(2\;{\rm GeV})=0.068\pm0.038$.
It is curious that the value of $a_2$ reported by \texttt{LPC} Collaboration is about twice greater than that reported by the \texttt{RQCD} Collaboration.
This discrepancy might be attributed to the fact that the LaMET approach receives large power correction in the endpoint region.
On the other hand, it is very challenging for the local operator matrix element approach~\cite{RQCD:2019osh} to compute the higher Gegenbauber moments,
thus difficult to reconstruct the whole profile of the LCDA.

\noindent{\color{blue}\it Phenomenological exploration.}
We use the three-loop evolution equation~\cite{Braun:2017cih,Strohmaier:thesis} to evolve each $a_n$ evaluated at 2 GeV by lattice simulation to any intended
scale $\mu$. We only retain those Gegenbauer moments with $n$ up to 6. We also use the package {\tt FAPT}~\cite{Bakulev:2012sm} to evaluate the running QCD coupling constant
to three-loop accuracy.

For the sake of comparison, we take the pion EM form factor data in the spacetime region from {\tt NA7} collaboration~\cite{NA7:1986vav}, {\tt Cornell} data compiled by Bebek {\it et al.}~\cite{Bebek:1977pe}, and the reanalyzed {\tt JLab} data~\cite{JeffersonLab:2008jve}, and take the time-like pion EM form factor data
entirely from the {\tt BaBar} experiment~\cite{BaBar:2012bdw}. We discard many irrelevant small-$Q^2$ data.

\begin{figure}[ht]
\centering
\includegraphics[width=0.235\textwidth]{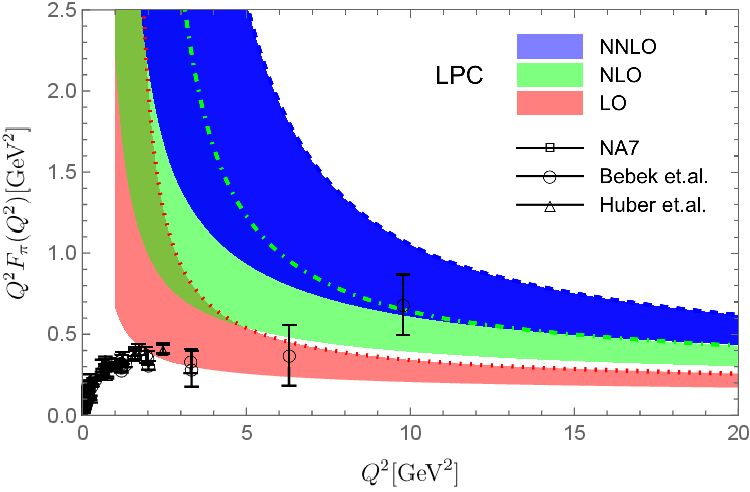}
\includegraphics[width=0.235\textwidth]{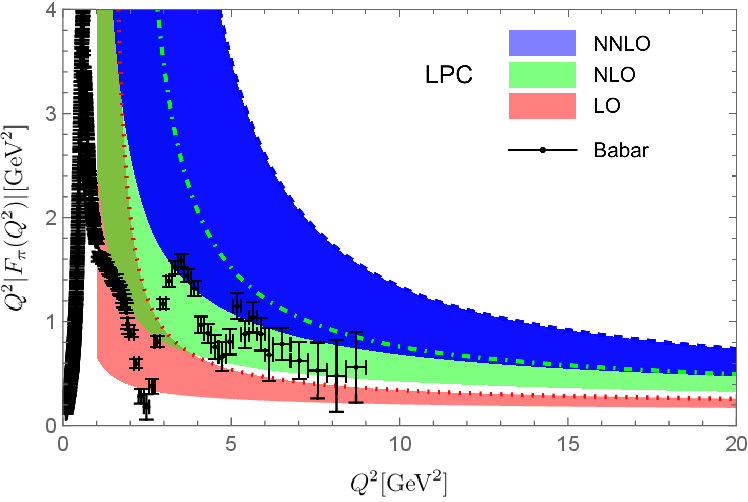}
\caption{Theoretical predictions vs. data for $Q^2 F_\pi(Q^2)$ in the space-like (left panel) and time-like (right panel) regions.
We take the central values of $a_2$, $a_4$ and $a_6$ determined by \texttt{LPC}.
The red, green and blue curves correspond to the LO, NLO and NNLO results, and the respective bands are obtained by sliding $\mu$ from $Q/2$ to $Q$.
Experimental data points are taken from {\tt NA7}~\cite{NA7:1986vav}, {\tt Bebek et~al.}~\cite{Bebek:1977pe}, {\tt Huber et al.}~\cite{JeffersonLab:2008jve} and
{\tt BaBar}~\cite{BaBar:2012bdw}.}
\label{plot:theory:vs:data:LPC}
\end{figure}

\begin{figure}[ht]
    \centering
    \includegraphics[width=0.235\textwidth]{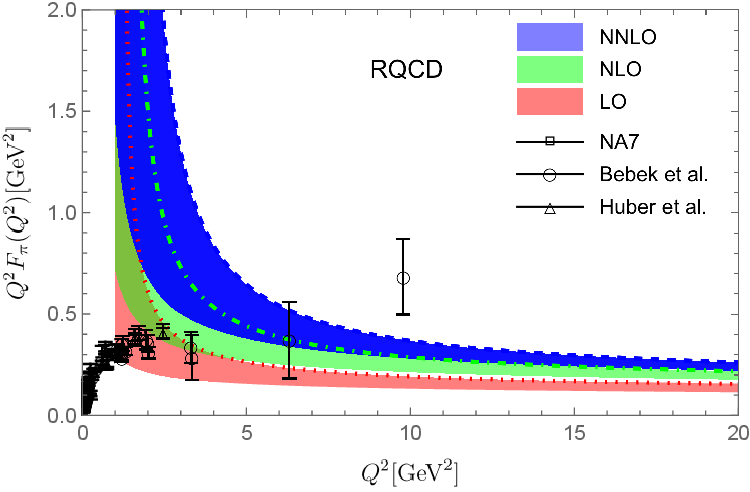}
    \includegraphics[width=0.235\textwidth]{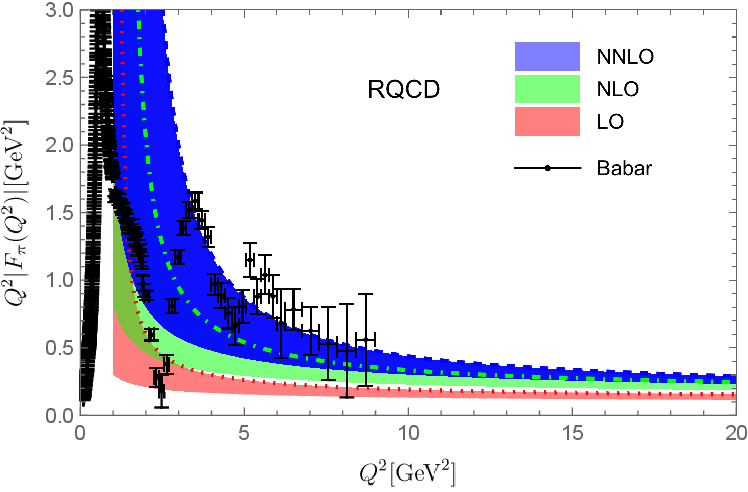}
    \caption{Same as Fig.~\ref{plot:theory:vs:data:LPC}, except the predictions are made by taking the central value of $a_2$
    determined by \texttt{RQCD}, with $a_4$ and $a_6$ set to zero.}
\label{plot:theory:vs:data:RQCD}
\end{figure}

In Fig.~\ref{plot:theory:vs:data:LPC} and Fig.~\ref{plot:theory:vs:data:RQCD}, we confront our predictions at various perturbative accuracy with the available data, including both space-like and
time-like regime.  One clearly visualizes that NNLO correction is positive and substantial.
In Fig.~\ref{plot:theory:vs:data:LPC}, we set the various Gegenbauer moments of pion LCDA to the central values given by \texttt{LPC} Collaboration~\cite{LatticeParton:2022zqc}. It appears the NNLO predictions are significantly overshooting the experimental data at large $Q^2$ ($>5\;{\rm GeV}^2$),
especially for the time-like regime with high statistics data. This symptom is mainly due to the large value of $a_2$.

In Fig.~\ref{plot:theory:vs:data:RQCD} we present our predictions with $a_2$ taken from \texttt{RQCD} while setting the values of $a_4$ and $a_6$ to zero.
We find satisfactory agreement between our NNLO predictions and the data,
both in space-like and time-like regimes. This may indicates that the value of $a_2$ given by \texttt{RQCD} might be more trustworthy.
It is of utmost importance for \texttt{RQCD} and \texttt{LPC} collaborations to settle the discrepancy in the value of $a_2$ in the future.

The prospective Electron-Ion Collider (EIC) program plans to measure the space-like pion EM form factor with $Q^2$ as large as $30\;{\rm GeV}^2$~\cite{Bylinkin:2022rxd},
where perturbative QCD should be definitely reliable. We are eagerly awaiting to confronting our NNLO predictions with the future \texttt{EIC} data.

\vspace{0.1 cm}

\noindent{\color{blue}\it Summary.} In this work we report the first calculation of the NNLO QCD corrections to the pion electromagnetic form factor.
We have explicitly verified the validity of the collinear factorization to two-loop order for this observable, and obtain
the UV, IR-finite two-loop hard-scattering kernel in closed form. The NNLO QCD correction turns to be positive and important.
We then confront our finest theoretical predictions with various space-like and time-like pion form factor data.
Our phenomenological study reveals that adopting the second Gegenbauer moment computed by \texttt{RQCD} can yield a decent agreement with large-$Q^2$ data (above the resonance
region in the time-like case).
Nevertheless, to make a definite conclusion, it seems imperative to resolve the discrepancy between \texttt{LPC} and \texttt{RQCD} Collaboration on the value of $a_2$
in the future study.
Furthermore, we look forward to the future high-statistics larger-$Q^2$ pion EM form factor data for critically testing our NNLO predictions.
It is also very interesting to confront our NNLO predictions with the available high-quality
kaon EM form factor data.

\begin{acknowledgments}
We are indebted to Jun Hua for providing the value of the Gegenbauer moment $a_6$ based on the recent LPC study~\cite{LatticeParton:2022zqc}.
The work of L.-B.~C. is supported by the NNSFC Grant No. 12175048.
The work of W.C. is supported by National Natural Science Foundation of China under contract No. 11975200.
The work of F.~F. is supported by the NNSFC Grant No. 12275353.
The work of Y.~J. is supported in part by the NNSFC Grants No.~11925506, No.~12070131001 (CRC110 by DFG and NSFC).

\vspace{0.2 cm}

{\bf Note added.} The published version of this work [L.~B.~Chen, W.~Chen, F.~Feng and Y.~Jia,
Phys. Rev. Lett. \textbf{132}, no.20, 201901 (2024)] entails a mistake.
The renormalized two-loop form factor $Q^2 F^{(2)}(u,v)$ 
in \eqref{two:loop:matching:equation} has missed some finite pieces, which  
lead to the incorrect expression of the two-loop hard-scattering kernel $T^{(2)}(x,y)$, 
and consequently the entries related to ${\cal T}^{(2)}_{mn}$ in Table~\ref{tab:Tmn-numeric-values} and Eq.~(18) 
need modification. This mistake is traced to the omission of some finite terms which arise from 
the lower order amplitudes upon conducting the QCD coupling constant renormalization under $\overline{\rm MS}$
scheme. We thank Jian Wang for informing us of the discrepancy between their unpublished work 
(Y.~Ji, B.-X.~Shi, J.~Wang, Y.~Wang, Y.-M.~Wang, H.-X.~Yu, to appear) and ours. 
After correcting this error, our results agree with theirs. Using the corrected form of the $T^{(2)}(x,y)$ turns out to 
have a minor phenomenological impact, {\it viz.}, the NNLO predictions in Fig.~2 and Fig.~3 remain almost intact.

\end{acknowledgments}

\appendix

\section{Asymptotical expressions of the hard-scattering kernel}

The analytic expressions of the hard-scattering kernels are too lengthy to be presented in the text.
Nevertheless, it is instructive to present their asymptotic expressions near the endpoint regimes.
For the one-loop hard-scattering kernel, we have
\begin{widetext}
\begin{subequations}
\bqa
\lim_{x \to 0 \atop y \to 0}T^{(1)}(x,y,\mu)
& =&-\frac{ e_d}{36 x y} \bigg[12\ln^2(x y)-18\ln(x y)-\pi^2+30-3(8\ln(x y)-3)L_{\mu}\bigg], \\
\lim_{x \to 0 \atop y \to 1} T^{(1)}(x,y)
 & =&-\frac{ e_d}{12 x}\bigg[4\ln^2 x-\ln x\ln\bar{y}-7\ln x-\ln\bar{y}+15-(8\ln x -3)L_{\mu}\bigg]\nn\\
 &+ &[x\to \bar{y}, \bar{y}\to x,e_d\to -e_u].
\eqa \label{asymptotic:T1:end:point:region}
\end{subequations}
\end{widetext}
The limiting behavior of $T^{(1)}(x,y,\mu)$ in two other corners, $x \to 1, y \to 1$ and $x \to 1, y \to 0$, can be obtained from the above formulas by
making the substitutions $x\to \bar{x},y \to \bar{y}$ and $e_u \leftrightarrow -e_d$.

For the two-loop hard-scattering kernel, we have
\begin{widetext}
\begin{subequations}
\bqa
\lim_{x \to 0 \atop y \to 0} T^{(2)}(x,y,\mu)
& =&-\frac{ e_d  }{18 x y}\bigg[\ln^4(x y)-\frac{15}{2}\ln^3(x y)-(\frac{5}{3}\pi^2-\frac{367}{8})\ln^2(x y)-\frac{81}{4}\ln x  \ln y  \nonumber\\
& &+(73\zeta_3+\frac{137}{48}\pi^2-\frac{1169}{8})\ln(x y)  +\frac{83}{60}\pi^4 - \frac{219}{2}\zeta_3-\frac{269}{24}\pi^2+\frac{3177}{16}\nonumber\\
& &+\frac{1}{8}(8\ln(x y)-3)(4\ln(x y)-15)L^2_{\mu}\nonumber\\
& &-(4\ln^3(x y)-21\ln^2(x y)-\frac{1}{3}(10\pi^2-\frac{537}{2})\ln(x y)+4\zeta_3+\frac{25}{4}\pi^2-81)L_{\mu}\bigg].
\\
\lim_{x \to 0 \atop y \to 1} T^{(2)}(x,y,\mu)
& =&-\frac{ e_d  }{18 x }\bigg[ \ln^4 x  -\frac{1}{2} \ln^3 x  \ln  \bar{y} -\frac{5}{32} \ln^2 x \ln^2 \bar{y}  -\frac{1}{6} \ln  x  \ln^3 \bar{y}  \nonumber\\
& &-8\ln^3 x +3 \ln^2 x \ln \bar{y}+\frac{19}{8} \ln x \ln^2\bar{y}-\frac{1}{6} \ln^3\bar{y}\nonumber\\
& &-\frac{1}{48} \left(\left(89 \pi ^2-2181\right) \ln^2 x+2 \left(225-14 \pi ^2\right) \ln x  \ln \bar{y}-165 \ln^2\bar{y}\right) \nonumber\\
& &- \frac{1}{8} \left(\left(-395 \zeta_3-47 \pi ^2+1106\right) \ln x+\left(29 \zeta_3 +\pi ^2+132\right) \ln \bar{y}\right)\nonumber\\
&&+\frac{1}{80} \left(32 \pi ^4-7670 \zeta_3-430 \pi ^2+11505\right)+\nonumber\\
& & +(4\ln^2 x-\frac{33}{2}\ln x-\frac{2}{3}\pi^2+\frac{13}{8})L^2_{\mu}- (4\ln^3 x-\ln^2 x\ln\bar{y}\nonumber\\
& &-\frac{1}{2}\ln^2\bar{y}\ln x-22\ln^2 x+\frac{15}{4}\ln x\ln\bar{y}-\frac{1}{2}\ln^2\bar{y}+(\frac{357}{4}-4\pi^2)\ln x\nonumber\\
& &+\frac{19}{4}\ln \bar{y}-7\zeta_3+\frac{14}{3}\pi^2-\frac{147}{4})L_{\mu}\bigg]\nonumber\\
 & &+[x\to \bar{y}, \bar{y}\to x,e_d\to -e_u].
\eqa
\label{asymptotic:T2:end:point:region}
\end{subequations}
\end{widetext}
Similar to the one-loop case, the limiting behavior of $T^{(2)}(x,y,\mu)$ in two other corners, $x \to 1, y \to 1$ and $x \to 1, y \to 0$,
can also be deduced by making the substitutions $x\to \bar{x},y \to \bar{y}$ and $e_u \leftrightarrow -e_d$.

Inspecting \eqref{asymptotic:T1:end:point:region} and \eqref{asymptotic:T2:end:point:region}, one observes that
$T^{(1)}(x,y,\mu)$ contains double endpoint logarithms exemplified by $\ln^2(x y)$ and $\ln^2 x$,
while $T^{(2)}(x,y,\mu)$ of entail quartic endpoint logarithms $\ln^4(x y)$ and $\ln^4 x$.
It is curious whether such endpoint logarithms can be resummed to all orders in $\alpha_s$ or not.

\begin{table}[ht]
\begin{tabular}{c|cc|ccc}
\hline\hline
(m,n) & $\tilde{c}_1$ & $\tilde{c}_2$ & $\tilde{d}_1$ & $\tilde{d}_2$ & $\tilde{d}_3$ \\
\hline\hline
(0,0) & 13.8194 & 39.4375 & 36.971 & 249.85 & 475.367 \\
(0,2) &  27.1946 & 94.1639 & 84.57 & 662.361 & 1417.37 \\
(0,4) & 29.925 & 118.728 & 103.512 & 902.131 & 2122.1 \\
(0,6) & 33.65 & 142.097 & 122.254 & 1125.79 & 2794.25 \\
(2,2) & 49.3167 & 204.84 & 176.877 & 1601.67 & 3877.72 \\
(2,4) & 52.2326 & 244.914 & 206.395 & 2067.27 & 5486.7 \\
(2,6) & 58.0254 & 286.907 & 239.323 & 2518.04 & 7020.33 \\
(4,4) & 54.25 & 284.119 & 234.524 & 2583.99 & 7485.88  \\
(4,6) & 59.879 & 328.872 & 269.162 & 3103.02 & 9408.85 \\
(6,6) & 65.949 & 378.914 & 307.704 & 3703.61 & 11726.6 \\
\hline\hline
\end{tabular}
\caption{The numerical values for ${\cal T}^{(1)}_{mn}\big|_{\rm asy} = (\tilde{c}_1 L_\mu + \tilde{c}_2)$ and ${\cal T}^{(2)}_{mn}\big|_{\rm asy} = (\tilde{d}_1 L_\mu^2 + \tilde{d}_2 L_\mu+\tilde{d}_3)$,
with $0\le m,n\le 6$, which are obtained by taking the asymptotical expressions for $T^{(1,2)}$ in \eqref{asymptotic:T1:end:point:region} and \eqref{asymptotic:T2:end:point:region}. }
\label{tab:Tmn-numeric-aym}
\end{table}

As an interesting exercise, we use the asymptotic expressions \eqref{asymptotic:T1:end:point:region} and \eqref{asymptotic:T2:end:point:region} to
evaluate the convolution integrals as defined in (17) of the main text. We divide the integrate regions into four parts:
$0<x,y<\frac{1}{2}$, $\frac{1}{2}<x,y<1$, as well as $(0<x<\frac{1}{2}, \frac{1}{2}<y<1)$ and $(\frac{1}{2}<x<1,0<y<\frac{1}{2})$.
It is found that the contributions from the last two regions cancel with each other,
and the first two regions yield identical results once the replacement $e_d\leftrightarrow -e_u$ is made.
For the sake of clarity, we also tabulate the numerical results  of ${\cal T}^{(k)}_{mn}\big|_{\rm asy}$ in Table~\ref{tab:Tmn-numeric-aym}.
In comparison with Table~I in the main text, we observe that the agreement between ${\cal T}^{(1,2)}_{mn}\big|_{\rm asy}$ and the exact results
becomes increasingly satisfactory as $m,n$ increase.


\end{document}